\documentclass[aip]{revtex4-1}
\usepackage{graphicx}
\usepackage{epstopdf}

\begin{document}

\title{The THETA - temperature of Interacting Self Avoiding Walk on Face Centered Cubic Lattice}

\author{Asweel Ahmed A. Jaleel$^1$, M. Ponmurugan$^2$ and S. V. M. Satyanarayana }
\email[]{svmsatya@gmail.com}

\affiliation{$^1$ Department of Physics, Pondicherry University, Kalapet,Pondicherry, 605014, India \\
$^2$ Department of Physics, School of Basic and Applied Sciences, \\
Central University of Tamilnadu, Tiruvarur 610 004, Tamilnadu, India.} 

\date{\today}

\begin{abstract}
 Interacting Self Avoiding Walk (ISAW) on a lattice is a simple model to study the
Coil to Globule transition of linear homopolymers. The temperature at which the transition
takes place is called the theta temperature. The value of theta temperature depends on the
chosen lattices. The value of theta temperature for ISAW on FCC lattice is reported as $6.46$ 
in a simulation study [J. Chem. Phys. {\bf 135}, 204903 (2011)]. Latter it has been
reported as $7.614$ [J. Chem. Phys. {\bf 138}, 024902 (2013)]. 
Simulations in these two studies involved long chains with over 2000 monomers.
In this paper, we present a method that gives a reliable estimate of the theta temperature of ISAW on FCC lattice 
using chains of shorter walk lengths (less than 100 monomers).
We compute density states of ISAW on FCC lattice by employing growth walk
algorithm and then we use the recently introduced  pseudo order parameter method
to estimate the theta temperature. The value obtained from our method agrees very
well with latter reported result of $7.614$. In order to corroborate our result we also estimate 
the theta temperature using the celebrated partition function zeroes method. 

\end{abstract}


\maketitle

\section{Introduction}
Conformational transition of a linear homo-polymer chain
from its coil state to globule state by a change in physical (temperature)
or chemical (solvent concentration) conditions in dilute solution has been a widely studied phenomena\cite{pgd}. This transition is
due to competition between excluded volume effect and the monomer - monomer interactions.
These interactions get balanced at a particular temperature called {\it theta temperature}
or at a particular solvent called {\it theta solvent} for which the linear polymer behaves as an
ideal chain \cite{pgd}. The ideal chain behaviour has  been studied by a random walk model of the polymer whose
ensemble average of radius of gyration $R_g$ scales with length of the chain $N$ as
$<R_g^2> \sim N^{2\nu}$, where $\nu$ is known as the Flory's exponent which is equal
to $1/2$ for the ideal chain \cite{van}.

The excluded volume interaction is included in random walk model through a constraint
that segments of walk do not intersect with each other. Such a model is called as Self Avoiding Walk which is useful 
to study polymer in  a good solvent condition or at high temperature \cite{van}. Interaction between the 
monomers can be an attractive interaction among non bonded nearest neighbours.  This is called as Interacting Self Avoiding Walk. The
continuous\cite{you,yon}, lattice\cite{pgd,flory1,flory2,pon1,pon2} and off lattice\cite{off1,off2} versions of this model have been studied extensively.

Interacting Self Avoiding Walk (ISAW) on a lattice is a simple model to study the coil to globule transition
of linear polymers. The transition temperature, known as theta temperature depends on the chosen
lattice \cite{bar}. Since the system behaves as an ideal chain at theta temperature, it can be estimated based on the
analysis of the value of Flory exponent $\nu$ evaluated from the scaling of ensemble average of the radius of gyration or end to end distance
with walk length $N$. That is the temperature at which $\nu=1/2$ has been identified as the theta temperature \cite{jir1}.
Several other methods can also be used to obtain theta temperature notably analysis of partition function zeroes \cite{jae1} and
inflection points\cite{inf1,inf2} of microcanonical entropy. In a recent study we have introduced a pseudo order parameter method to identify the theta temperature \cite{svm}.

Theta temperature of ISAW has been obtained for different types two and three dimensional lattices\cite{pon2,svm,gras,gras1,ramp,jae,wit,yon,jir2} . In particular, recent Monte Carlo simulation of ISAW on Face Centered Cubic (FCC)lattice reports an estimate of theta temperature as $6.46$ \cite{jir1}. The theta temperature reported in another work from same group is $7.614$ \cite{jir2}. These studies employed Metropolis Monte Carlo method  to compute Flory exponent as a function of temperature. The theta temperature is estimated as the value for which the Flory exponent is $1/2$.  Two different values of theta temperature reported on the same lattice initiated us to study ISAW on FCC lattice systematically. We compute density of states of ISAW on FCC lattice for the first time. We estimate theta temperature of ISAW on FCC lattice using recently proposed pseudo order parameter method and well known analysis of partition function zeros on complex temperature plane. We find that a reliable value for theta temperature can be estimated from pseudo order parameter method using even ISAW of short walk lengths.

The paper is organized as follows. Computation of density of states and microcanonical entropy of ISAW on FCC lattice for different walk lengths using the concept of atmosphere is presented in section II. Estimation of theta temperature using pseudo order parameter method \cite{svm} as well as analysis of partition function zeros \cite{yang,lie,lee,jae,jae1} is discussed in section III. Conclusions are presented in section IV. 

\section{Density of states estimates of ISAWs on FCC lattice}

In our simulation we employed growth walk algorithm \cite{baumg,zal,rr} to generate the self avoiding
walk on a FCC lattice. There are many growth walk algorithms that can be used to compute density of states of
ISAWs \cite{pon1,pon3}. In particular, the algorithm we have employed to generate SAWs are
called Rosenbluth-Rosenbluth (RR) walk or Kinetic growth walk \cite{rr}.
In RR algorithm, the walk start at arbitrary chosen origin on a lattice and search for the unoccupied nearest neighbors sites.
The walk will randomly move to any one of the available unoccupied nearest neighbour site with equal probability.
This process continues until  the walk reaches a given walk length $N$. In walks, if there is no unoccupied nearest neighbour
sites available for a walk to proceed further to reach given $N$, such walks are called trapped walks. These walks can be
discarded and a fresh walk can be started from the origin. All the SAWs generated in this algorithm are rooted at the origin of the lattice. By assiging an energy  $\epsilon$ to all non bonded nearest neighbors (nbNN) of succesfully generated self avoiding walk of length $N$ one can  obtain ISAWs. The energy of a configuration with $`m'$
nbNN contacts are $E = m \epsilon$. Without loss of generality one can take $\epsilon=-1$ for attractive interaction.
An ensemble of such walks are generated.

\begin{figure}[h]
\includegraphics[width=0.8\textwidth]{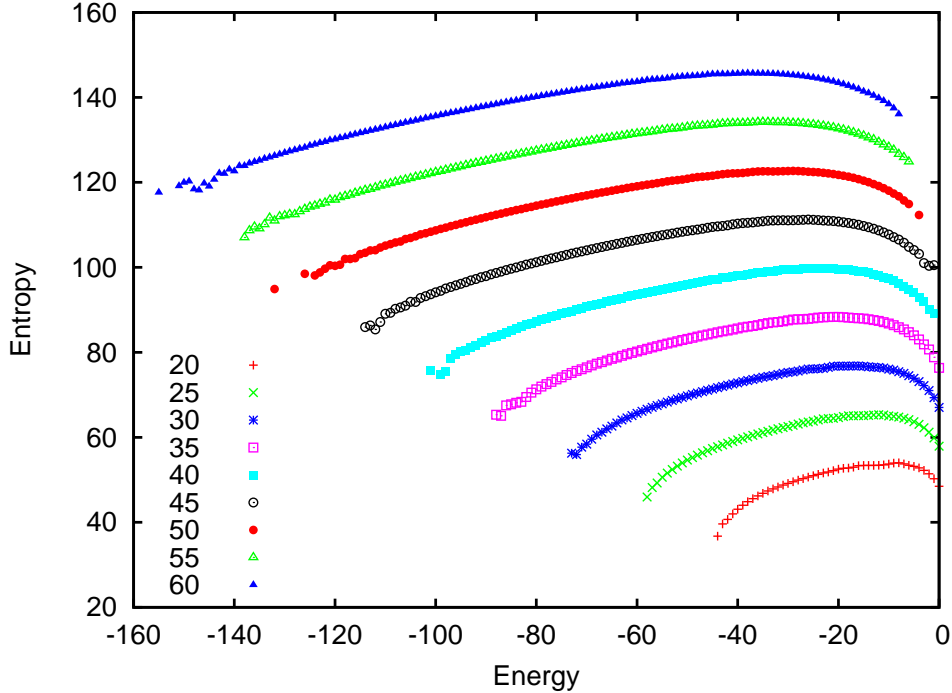}
\caption{Microcanonical entropy of ISAWs on FCC lattice for different walk lengths}
  \label{fig:entropy}
\end{figure}

Statistical mechanics of a given system can described, once we know the Density of states (DoS). Counting Self Avoiding Walk conformations corresponding to a particular energy  will in turn give the DoS for that particular energy. We used the concept of atmosphere in our growth algorithm and counted the DoS of ISAWs \cite{rech,prel,pon1}. For a particular conformation $\mathcal{C}$,
let there be $a_i$ number of unoccupied nearest neighbors sited available at $i^{th}$ step, then $a_i$ is the local
atmosphere for that step. Global atmosphere or simply the 'atmosphere' $\mathcal{A}$ for the successfully generated $N$ step
walk is defined as \cite{prel,pon1},
\begin{equation}
  A_N(\mathcal{C}) = \prod _{i=1}^N a_i (\mathcal{C}).
\end{equation}
The atmosphere for trapped walk is always zero \cite{pon1}. Since FCC lattice has $12$ nearest neighbors, highest number of nearest
neighbors $z$ in any other lattice in three dimension, the trapped conformations are  considerably less in the FCC lattice.
Let there are $m(\mathcal{C}$) number of nbNN contacts in a particular conformation $\mathcal{C}$, energy of that particular conformation $E(\mathcal{C})$ is $m\epsilon$ or $-m$. Let $C_N^m$ denotes the Density of states (DoS) ISAWs of length $N$ having energy -m,
the Monte Carlo estimate of DoS can be obtained from the atmosphere as \cite{prel,pon1}
\begin{equation}
  C_{N,m}^{est} = \frac{1}{M}\sum_{l=1}^{M_m} A_N(\mathcal{C}_l)
\end{equation}
Here $M$ is the total number of walks including trapped configurations. $M_m$ is the total number of walks with energy $-m$.
The monte carlo estimate of microcanonical entropy can be found from the estimated density of states as,
\begin{equation}
   S_N (E=-m) = log(C_{N,m}^{est})
\end{equation}
Figure ~\ref{fig:entropy} presents the computed microcanonical entropy for ISAWs of different walk lengths.
We have computed DoS up to walk length $300$. In a small cluster, for $300$ step walk, the computation took more than a month to complete generation of $10^9$ walks. Hence we limited our further analysis for ISAW of walk lengths less than $100$. We show that the theta temperature can be reliably estimated using recently proposed pseudo order parameter method using ISAW of short walk lengths. We observe however higher walk lengths up to $300$ are required for a reliable estimate of theta temperature using the analysis of partition function zeros.

\section{Estimation of theta temperature of ISAWs on FCC lattice}
 At a given inverse temperature $\beta =\frac{1}{k_BT}$ where $k_B$ is the Boltzmann constant (which is taken as unity),
the Canonical Partition function can be calculated from the  DoS estimates as
 \begin{equation} 
   C_N (\beta) =  \sum_{m=0} ^{m_N^{max}} C_{N,m} \mbox{exp}[-m \epsilon \beta],  \label{cpf}
 \end{equation}
where $m_N^{max}$ is the maximum number of contacts that an N step walk can have.
The canonical probability distribution is defined as \cite{svm},
\begin{equation}
    P_N(\beta,m) = \frac{C_{N,m} \mbox{exp}[-m\epsilon\beta]}{C_N (\beta)}.
\end{equation}

\begin{figure}[h]
\includegraphics[width=0.8\textwidth]{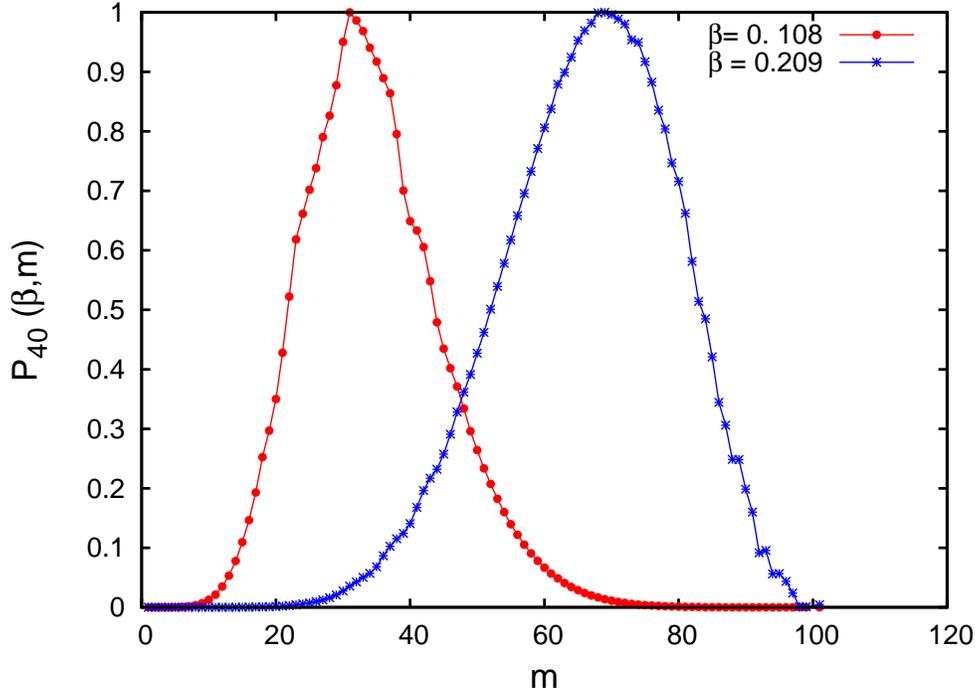}
\caption{Canonical probability distribution at two different $\beta$ below and above $\beta_{\Theta}$ for ISAW of walk length 40.}
  \label{fig:cpd}
\end{figure}

It may be noted that in the present work, $\epsilon=-1$. It is well established that the coil-globule transition in polymers are second order transitions \cite{pgd,klu,owc}. Canonical probability distribution as a function of number of contacts is shown in fig.~\ref{fig:cpd} for two values of inverse temperatures $\beta$ on either side of  inverse theta ($\beta_\Theta$) temperature. For second order transitions canonical probability distribution exhibits a single peak with the location of maximum of the peak varying continuously with $\beta$ across $\beta_\Theta$. In contrast, the canonical probability distribution shows a double hump for systems exhibiting first order phase transition and location of maximum of the distribution discontinuously jump from one peak to the other at transition temperature. This jump can be used to estimate the transition temperature  in first order transitions.

\subsection{Pseudo order parameter method}

It has been observed recently that the canonical probability distribution for systems exhibiting second order transition shows a single peak with  an asymetric long tail \cite{svm}. The long tail jumps from one side to other side across the transition. Based on this observation a pseudo order parameter is defined to obtain a binary signal of theta point transition on two dimensional square lattice \cite{svm}. For ISAWs on FCC lattice, it can be seen from fig.~\ref{fig:cpd} that long tail is on higher contact side for $\beta = 0.108 < \beta_\Theta$ and it is on the low contact side $\beta = 0.209 > \beta_\Theta$.

\begin{figure}[t]
\includegraphics[width=0.8\textwidth]{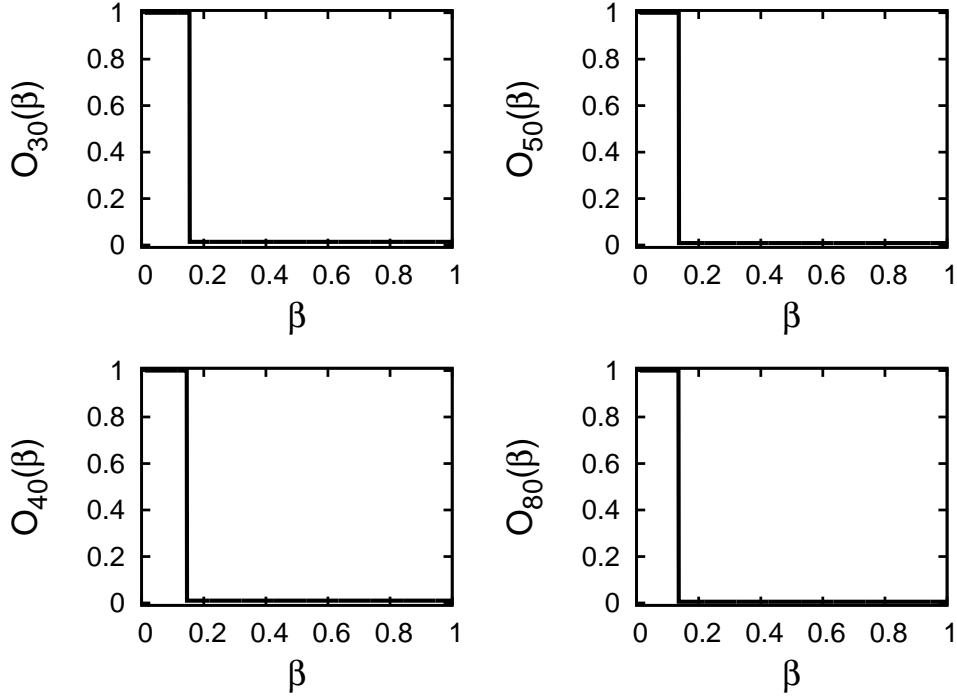}
\caption{Pseudo order parameter versus inverse temperature for different walk lengths}
  \label{fig:op}
\end{figure}

The pseudo order parameter defined based on the shifting of the tail of canonical probability distribution function
across the theta temperature transition is given by
\begin{equation}
    O_N (\beta) = \frac{m\big( min[P_N(\beta,m)]\big) }{m_N^{max}}
\end{equation}
	
\begin{figure}[t]
\includegraphics[width=0.8\textwidth]{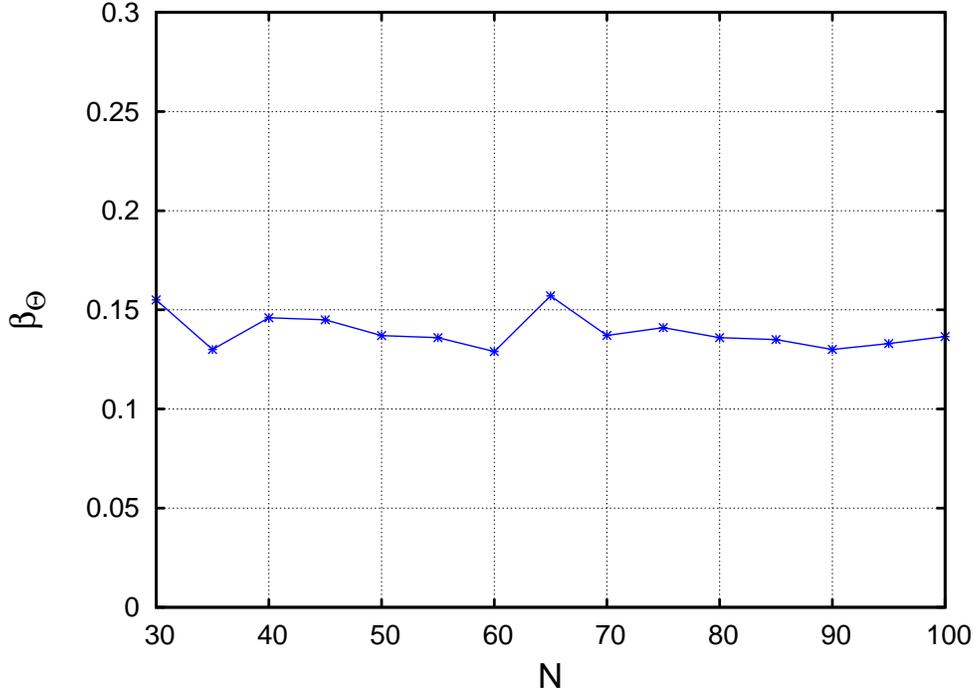}
\caption{Fluctuation of theta temperature for different walk lengths.}
  \label{fig:opf}
\end{figure}

Figure ~\ref{fig:op} shows the pseudo order parameter verses inverse temperature for different walk lengths. It can be seen that the order parameter registers a jump from $0$ to $1$ and results in an unmistakable location and estimation of theta temperature. The pseudo order parameter is 0 for large value of $\beta$ (small value of temperature). This means the minimum of canonical probability distribution occurs for lowest value of contacts, which implies that the conformations with low value of contact is least possible. Thus the polymer chain modeled by ISAW is in a globule phase, where the number of contacts is large. On the other hand, for small value of $\beta$ (high temperature) the pseudo order parameter is 1, implying the high contact configuration least posible. This corresponds to coil phase.

It has been observed that the inverse theta temperature corresponding to different walk lengths fluctuates with $N$. These fluctuations are observed to be either bound or diminish with increasing $N$ (fig. ~\ref{fig:opf} ). As compared to earlier study \cite{svm}, we find that such a fluctuation is small on FCC lattice because large nearest neighbour sites. This fluctuation is diminished further even for walk length below $100$. Thus we limited our analysis for walk length up to $100$ as mentioned earlier. The inverse temperature corresponding to the theta transition estimated using pseudo order parameter is $\beta_\Theta$ = 0.1385 $\pm$ 0.0085. This is in agreement with earlier reported result \cite{jir2} of $T_\theta= 7.614$ which in terms of inverse temperature is $\beta_\Theta = 0.1313$.  Merit of the pseudo order parameter method
on FCC lattice is that an accurate estimate of theta temperature can be obtained from ISAWs of relatively small walk lengths. In order
to corroborate our result we also used partition function zeroes method to identify theta temperature.

\subsection{Partition Function Zeros}
  In the thermodynamic limit, Yang and Lee proposed a general theory of phase transition based on the zeros of grand canonical partition function on complex fugacity plane, which they illustrated for Ising model in external field \cite{yang,lie}. Fisher extended this approach to the zeros of Canonical partition function \cite{fish}. Peter Borrman et al., showed that it is indeed true for systems of finite size \cite{borr}. Partition zeros has been used in the study of conformational statistics of coil to globule transition of linear  homopolymers \cite{rap1,lee,jae,pon1,jae1,mark}. In the thermodynamic limit locus of zeros forms a unit circle and crosses the Real axis. The zero which lies on positive real axis give the phase transition temperature. But for finite system they never cross real-axis. First zero which is close to positive real-axis in upper half plane becomes important, and describes the phase transition temperature.
	
\begin{figure}[t]
\includegraphics[width=0.8\textwidth]{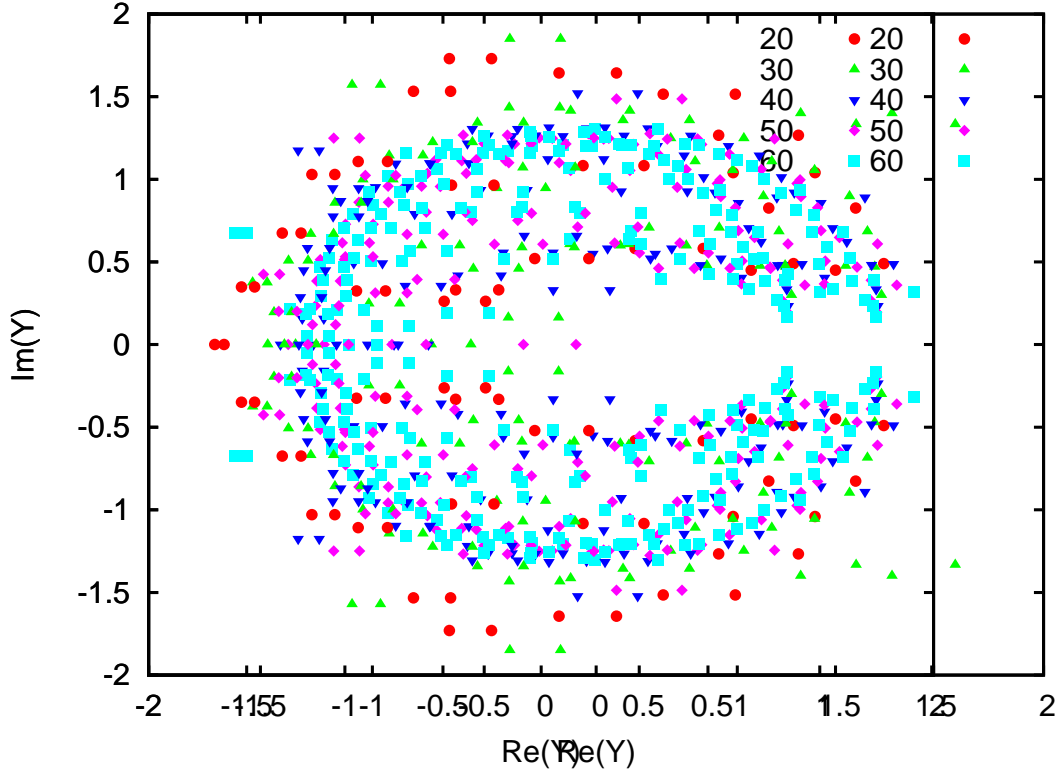}
\caption{Partition function zeros for different walk lengths on the positive quadrant of complex $Y$ plane.}
  \label{fig:pfz}
\end{figure}

	The canonical partition function  at a particular inverse temperature $\beta $ is defined in Equation ~\ref{cpf},
which can be recast as
	  \begin{equation}
	    C_N (\beta) = \sum _{m=0} ^{m_N^{max}} C_{N,m}y^m
	  \end{equation}
where y = exp($\beta$). Partition function is thus a polynomial of order $m_N^{max}$ and will have same number of roots which are complex. Figure ~\ref{fig:pfz} shows roots of the partition function zeros in the first quadrant of the complex $y$ plane for various walk lengths.

\begin{figure}[t]
\includegraphics[width=0.8\textwidth]{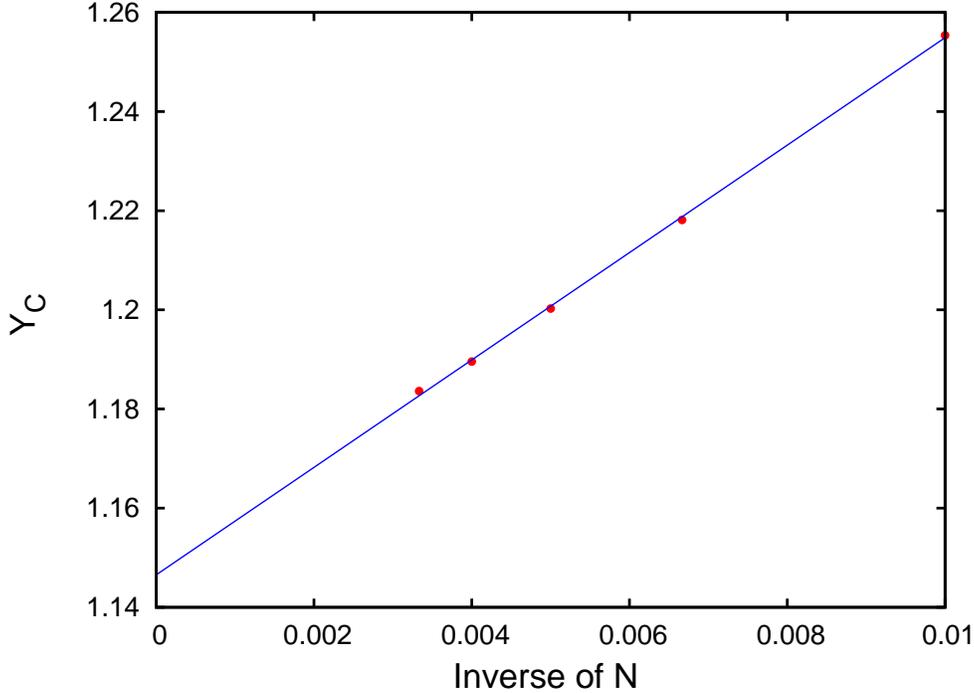}
\caption{Extrapolation of the real part of first zeros of partition function in the limit $N \to \infty$.}
  \label{fig:scale}
\end{figure}

The first zeros does not give us the true value of phase transition temperature because of finite size effects. So we have to do extrapolation of first zeros for longer walk lengths. In order to get a better approximation, we have used slightly bigger walk lengths namely 100 to 300. The real part of the first zeros as a function of 1/N is shown in fig.~\ref{fig:scale}. Let $Y_C$ be the real part of the first zero in the limit N $\rightarrow \infty$ . The value of $\beta_\Theta$ = $log Y_C$ =  $0.1367 \pm 0.0049$ computed in this method is in agreement with the value obtained using pseudo order parameter as well as with earlier reported result \cite{jir2}.

\section{Conclusion}
Using the notion of atmosphere of polymer growth algorithm
we computed the density of ISAW on FCC lattice. We estimated
the theta temperature of ISAW on FCC lattice by using pseudo order parameter
method. The theta temperature obtained from pseudo order parameter method
fluctuates around the true value for different walk lengths.
Compared to other lattices (where the coodination
number $z$ is less) we find that such a fluctuation is small for FCC lattice
(for which $z$ is large). Thus, our pseudo order parameter method provides a reliable
estimate of theta temperature even for shorter walk lengths. Our result showed that
the theta temperature of ISAW on FCC lattice obtained from our method agrees very well
with the earlier reported result of $7.614$. The same value obtained
from Partition function zeroes provides further support to our theta temperature
estimate.

\acknowledgments
Asweel Ahmed A Jaleel acknowledge UGC for grant (201112-MUS-KER-4227) under the scheme of Moulana Azad National Fellowship.

\bibliography{arbib}{}
\bibliographystyle{plain}
\end{document}